\documentclass[aps,prb,reprint,superscriptaddress,longbibliography]{revtex4-2}

\usepackage[switch]{lineno}
\usepackage{graphicx}
\usepackage{multirow}
\usepackage{amsmath,amssymb,amsfonts}
\usepackage{mathrsfs}
\usepackage{xcolor}
\usepackage{textcomp}
\usepackage{booktabs}
\usepackage{float}

\begin{document}

\title{Hybridization-controlled re-entrant electronic phase switching and moir\'e-confined states in twisted bilayer PtTe$_2$}

\author{Seoung-Hun Kang}
\thanks{These authors contributed equally to this work.}
\email{shkang@kisti.re.kr}
\affiliation{Department of Physics, Kyung Hee University, Seoul, Korea}
\affiliation{Department of Information Display, Kyung Hee University, Seoul, Korea}
\affiliation{Research Center for Technology Commercialization, Korea Institute of Science and Technology Information (KISTI), Seoul, Korea}

\author{Jeonghwan Ahn}
\thanks{These authors contributed equally to this work.}
\affiliation{The Anthony J Leggett Institute for Condensed Matter Theory, Department of Physics, University of Illinois at Urbana-Champaign, Urbana, Illinois, USA}

\author{Young-Kyun Kwon}
\email{ykkwon@khu.ac.kr}
\affiliation{Department of Physics, Kyung Hee University, Seoul, Korea}
\affiliation{Department of Information Display, Kyung Hee University, Seoul, Korea}

\begin{abstract}

Twisting a van der Waals bilayer changes not only the moir\'e periodicity but also the local stacking and interlayer hybridization. Here, we show, using fully relaxed first-principles calculations including spin--orbit coupling, band unfolding, and Brillouin-zone-integrated densities of states, that bilayer PtTe$_2$ exhibits a non-monotonic evolution between gapless and gapped electronic regimes. The $7.34^\circ$ structure remains gapless, whereas finite direct gaps appear at the sampled intermediate angles. The gap closes at the sampled $60^\circ$ configuration and reopens at higher angles. The direct gap shows an overall increase with the minimum local interlayer Pt--Pt separation, although the complete distribution of local stacking environments is required to account for deviations from this trend. At $7.34^\circ$, the low-energy states are concentrated predominantly in the AA-like regions of the otherwise gapless moir\'e cell. Controlled interlayer-separation scans show that increasing the layer spacing removes the near-$E_F$ crossings and opens a gap, consistent with weakened interlayer Te-$p_z$ hybridization. These results identify the redistribution of interlayer hybridization as the microscopic origin of the re-entrant gap evolution in twisted bilayer PtTe$_2$.
\end{abstract}


\maketitle

\section{Introduction}

Twisted bilayers of two-dimensional crystals form long-period moir\'e superlattices while simultaneously reorganizing local stacking, atomic relaxation, and interlayer coupling across the moir\'e cell\cite{Ponomarenko2013,Dean2013,Yankowitz2012,Ni2015}. This coupling between momentum-space reconstruction and real-space inhomogeneity underlies much of modern moir\'e physics. In graphene, it produces narrow bands and correlated phases\cite{Bistritzer2011,Cao2018CI,Cao2018SC}. In semiconducting transition-metal dichalcogenides, it has enabled strongly correlated and topological states, including integer and fractional Chern insulating phases in twisted MoTe$_2$\cite{Mak2022,Liu2021MoireCSR,Cai2023,Zeng2023,Park2023}. These developments have established that the electronic structure of a moir\'e material is governed not only by the superlattice period but also by the local registry and the orbital hybridization allowed by that registry\cite{Carr2020}.

Most studies have focused on graphene and semiconducting dichalcogenides, where band flattening, Coulomb interactions, and excitonic effects set the relevant low-energy scales. Metallic and semimetallic van der Waals bilayers provide a complementary limit. In these systems, twist can modify pre-existing band crossings by redistributing the interlayer hybridization, thereby offering a route to single-particle electronic phase control without requiring a correlation-driven insulating state.

PtTe$_2$ is particularly suitable for examining this mechanism. Bulk PtTe$_2$ hosts type-II Dirac fermions in the presence of strong spin--orbit coupling\cite{Yan2017}, while ultrathin PtTe$_2$ exhibits a pronounced dimensional crossover: the monolayer is semiconducting, while the bilayer is semimetallic\cite{Lin2020,Zhang2022}. This thickness dependence originates from the strong sensitivity of the electronic structure near $E_F$ to interlayer coupling, especially to the out-of-plane Te-$p_z$ component.

In a twisted PtTe$_2$ bilayer, the interlayer coupling cannot be represented by a single uniform parameter. AA-like, AB-like, and AC-like local registries coexist within one moir\'e cell, and each registry favors a distinct interlayer separation and orbital overlap. Similar registry-dependent reconstruction occurs in marginally twisted metallic dichalcogenide bilayers\cite{McHugh2023}. In PtTe$_2$, such structural inhomogeneity is expected to have a particularly strong electronic consequence because modest changes in the interlayer separation can determine whether the near-$E_F$ bands overlap or separate. The electronic phase should therefore be related to the spatial distribution of local interlayer environments rather than to the nominal twist angle alone\cite{Moon2014,Krongchon2023}.

Our previous diffusion Monte Carlo benchmark showed that r$^2$SCAN+rVV10 provides the closest overall description of the stacking-dependent binding energies and equilibrium interlayer separations of bilayer PtTe$_2$. It further showed that differences in the relaxed geometry can change whether the $21.79^\circ$ twisted structure is classified as metallic or gapped\cite{Kang2024PRR}. Recent PBE-D3 calculations reported twist-induced gap opening in large-angle NiS$_2$ and PtTe$_2$ moir\'e structures, including a gapped $60^\circ$ PtTe$_2$ configuration, and attributed the transition to weakened interlayer $d$--$p$ hybridization\cite{Li2025CPB}. However, the electronic evolution across a broad set of commensurate PtTe$_2$ structures, the recovery of a gapless state at a sampled intermediate-angle configuration, and the real-space distribution of low-energy states at small angles remain unresolved.

Here, we examine fully relaxed commensurate twisted bilayer PtTe$_2$ structures between $0^\circ$ and $90^\circ$. We find a non-monotonic sequence of gapless and gapped electronic regimes. The $7.34^\circ$ structure remains gapless but exhibits a pronounced domain dependence of the near-$E_F$ states. Finite direct gaps appear at the sampled intermediate angles; the gap closes at the sampled $60^\circ$ configuration and reopens at higher angles. By combining unfolded spectra, Brillouin-zone-integrated densities of states, local-registry analysis, partial charge densities, and controlled interlayer-separation scans, we identify separation-dependent Te-$p_z$ hybridization as the microscopic origin of this non-monotonic gap evolution.

\section{Results}

\subsection{Relaxed moir\'e structures and energetics of commensurate twist angles in bilayer PtTe$_2$}\label{subsec:struct}

\begin{figure}[t]
  \centering
  \includegraphics[width=0.98\linewidth]{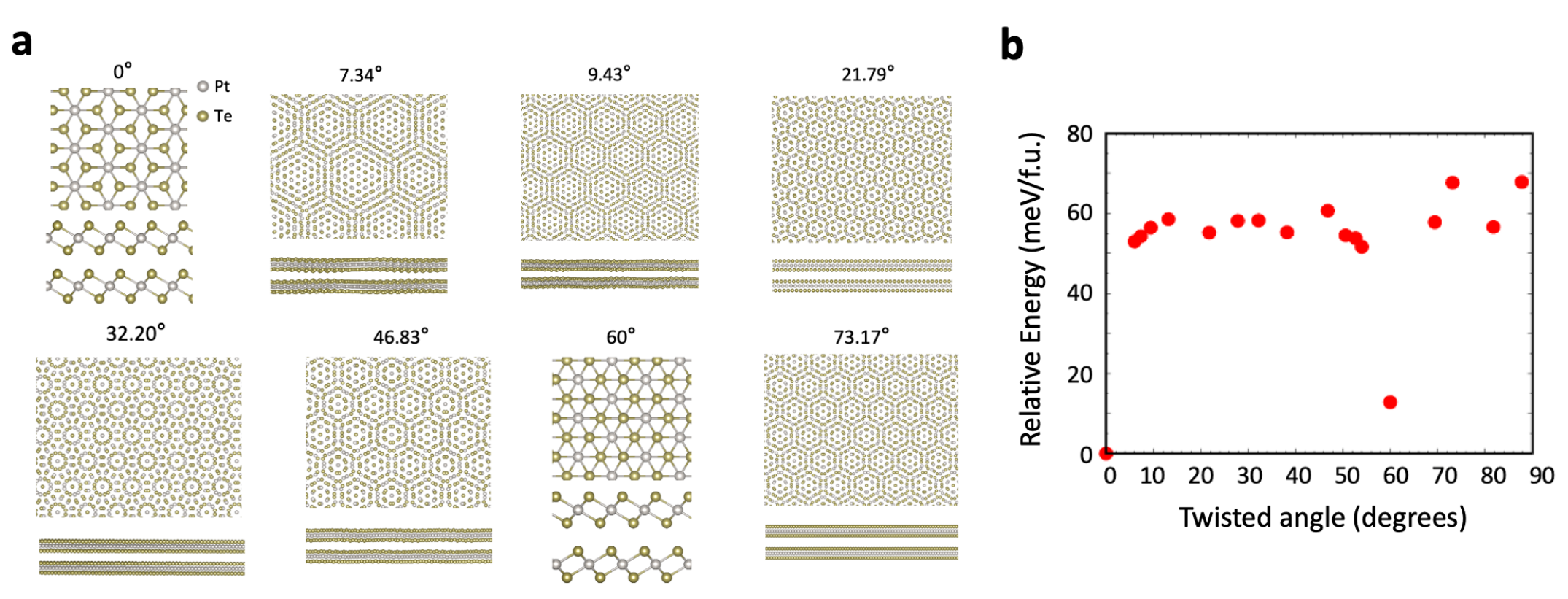}
  \caption{\textbf{Twisted bilayer PtTe$_2$ moir\'e structures and energetics.}
  \textbf{a,} Relaxed atomic structures at representative commensurate twist angles of $0^\circ$, $7.34^\circ$, $9.43^\circ$, $21.79^\circ$, $32.20^\circ$, $46.83^\circ$, $60^\circ$ and $73.17^\circ$. Top and side views are shown for each angle. \textbf{b,} Relative total energy per formula unit as a function of twist angle, referenced to the lowest-energy structure within the calculated commensurate-angle set. All structures are fully relaxed at each commensurate angle. The discrete energies compare the sampled commensurate models and
do not represent a continuous twist-angle energy landscape.}
  \label{fig1}
\end{figure}

We first define the structural set used throughout this work. To study the evolution of electronic states in a controlled manner, we constructed a series of commensurate twisted bilayer PtTe$_2$ supercells spanning $0^\circ$--$90^\circ$ and fully relaxed the atomic positions at each twist angle. Representative relaxed moir\'e structures are shown in the top and side views of Fig.~\ref{fig1}a. These structures establish the real-space stacking pattern that provides the structural basis for all electronic results discussed in the following.

The relaxed structures show that twisting is not simply a rigid rotation of one layer relative to the other. Rather, each commensurate angle produces a distinct moir\'e environment with its own distribution of local stacking registries and interlayer geometry. Because the interlayer interaction in PtTe$_2$ contains both non-directional van der Waals attraction and directional orbital hybridization, these structural changes are expected to directly reshape the low-energy electronic states.

Figure~\ref{fig1}b compares the relative total energies per formula unit of the sampled commensurate structures. These values document the energetic differences among the calculated structural models. Because the structures have different commensurate supercells and local stacking distributions, the discrete data should not be interpreted as a continuous twist-angle energy landscape or as an equilibrium preference for a particular twist angle.

\subsection{Non-monotonic electronic phase sequence in the unfolded spectra}\label{subsec:phase}

\begin{figure}[t]
  \centering
  \includegraphics[width=0.98\linewidth]{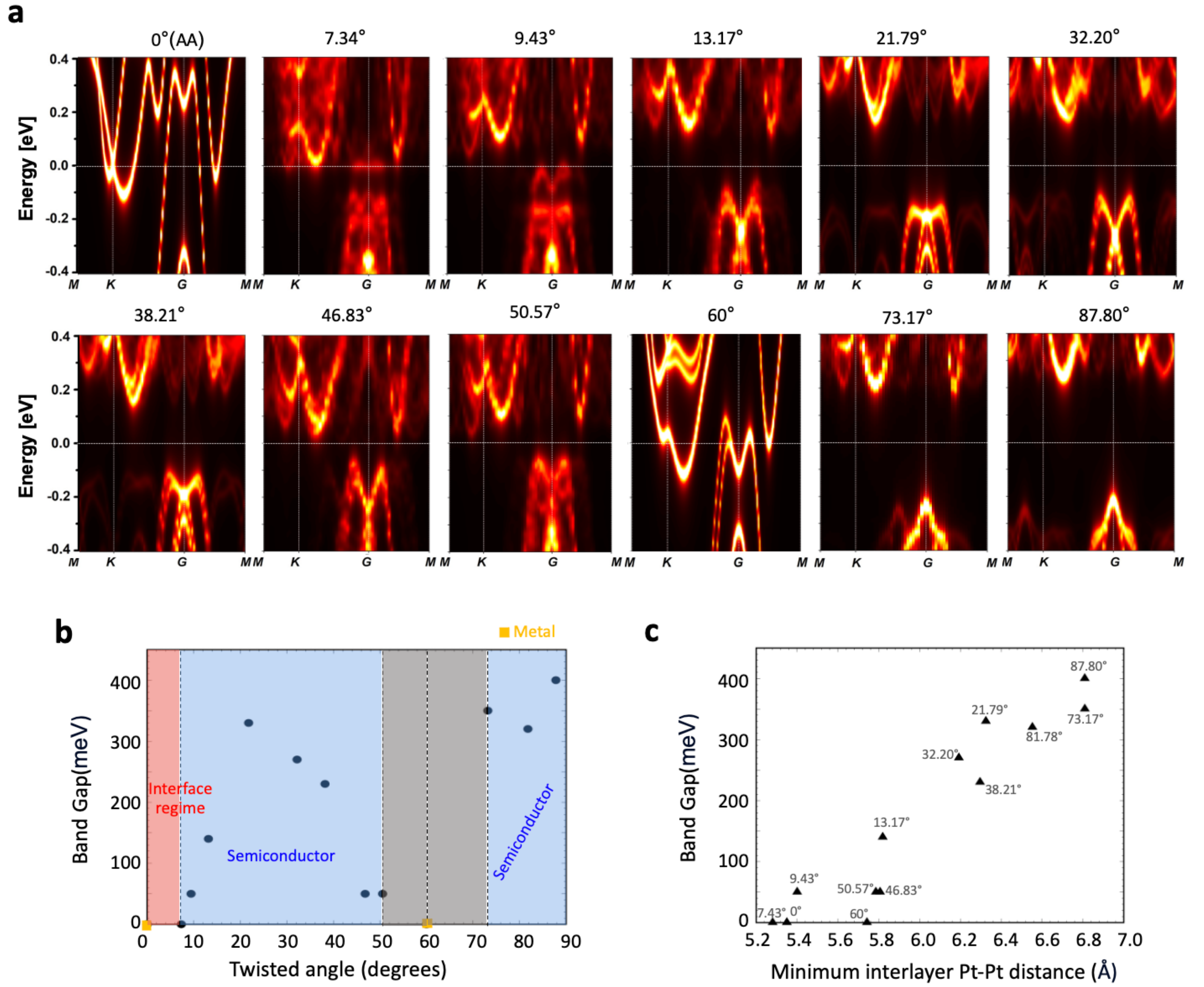}
  \caption{\textbf{Non-monotonic electronic phase sequence from unfolded spectra.}
  \textbf{a,} Unfolded spectral maps for commensurate twisted bilayer PtTe$_2$ at $\theta=0^\circ$ (AA), $7.34^\circ$, $9.43^\circ$, $13.17^\circ$, $21.79^\circ$, $32.20^\circ$, $38.21^\circ$, $46.83^\circ$, $50.57^\circ$, $60^\circ$, $73.17^\circ$, and $87.80^\circ$. The path is $M$--$K$--$\Gamma$--$M$. \textbf{b,} Direct spectral gap $E_g^{\mathrm{dir}}$ extracted from the unfolded spectra. The $7.34^\circ$ structure is assigned $E_g^{\mathrm{dir}}=0$ because appreciable spectral weight crosses $E_F$. The background shading is a guide to the eye for the electronic regimes represented by the sampled commensurate structures and does not denote precisely determined phase boundaries. \textbf{c,} $E_g^{\mathrm{dir}}$ plotted against the minimum local interlayer Pt--Pt separation. The $81.78^\circ$ point is included in \textbf{b} and \textbf{c} as an additional high-angle commensurate structure; its unfolded spectral map is omitted from \textbf{a} for compactness. The electronic classification is independently checked using the Brillouin-zone-integrated total densities of states shown in Supplementary Fig.~S1.}
  \label{fig2}
\end{figure}

We next examine the evolution of the low-energy electronic structure. Figure~\ref{fig2}a shows unfolded spectral functions along the primitive-cell path $M$--$K$--$\Gamma$--$M$ for the commensurate twist angles between $0^\circ$ and $90^\circ$. The spectral evolution is non-monotonic. Appreciable spectral weight crosses the Fermi level in the untwisted AA bilayer and in the $7.34^\circ$ structure. Then, finite direct spectral gaps appear at the sampled intermediate angles. The gap closes again for the sampled $60^\circ$ configuration before reopening in the high-angle structures.

Throughout this work, the electronic regime is assigned by combining the unfolded spectra with the Brillouin-zone-integrated total density of states. A structure is classified as gapless when states with appreciable unfolded spectral weight cross $E_F$ and the total density of states remains finite at $E_F$. A structure is classified as gapped when a finite direct spectral gap, $E_g^{\mathrm{dir}}>0$, is resolved in the unfolded spectrum and the total density of states is suppressed to the numerical background level over a finite energy interval around $E_F$ (Supplementary Fig.~S1). The values reported in Fig.~\ref{fig2}b are direct gaps extracted from the unfolded spectra, whereas the total density of states provides an independent Brillouin-zone-integrated consistency check of the electronic classification.

The extracted direct gaps are summarized in Fig.~\ref{fig2}b. The $7.34^\circ$ structure is gapless, and we therefore assign $E_g^{\mathrm{dir}}=0$. A small direct gap of approximately $0.06$~eV first appears at $9.43^\circ$. The gap increases to approximately $0.14$~eV at $13.17^\circ$ and reaches $0.24$--$0.33$~eV between the sampled $21.79^\circ$ and $38.21^\circ$ structures. It decreases to approximately $0.05$ and $0.06$~eV at $46.83^\circ$ and $50.57^\circ$, respectively, and closes at the sampled $60^\circ$ configuration. The gap reopens at higher sampled angles, reaching approximately $0.36$, $0.33$, and $0.40$~eV at $73.17^\circ$, $81.78^\circ$, and $87.80^\circ$, respectively.

Figure~\ref{fig2}c compares $E_g^{\mathrm{dir}}$ with the minimum local interlayer Pt--Pt separation, used here as a measure of the local layer spacing. The direct gap shows an overall increase with this structural quantity. The deviations from a one-parameter trend indicate that the shortest separation alone does not determine the electronic structure. The complete distribution of local separations and stacking environments also contributes to interlayer hybridization near $E_F$. Consistently, the total density of states is finite at $E_F$ for the gapless configurations and vanishes around $E_F$ for the gapped configurations within numerical resolution, while the orbital-projected densities of states show substantial Pt-$d$ and Te-$p$ character near the band edges (Supplementary Fig.~S1).

\subsection{Domain-selective low-energy weight in the small-angle moir\'e structure}\label{subsec:interface}

The gapless $7.34^\circ$ structure is electronically inhomogeneous in real space. Its relaxed moir\'e cell contains AA-like, AB-like, and AC-like local registries, which we identify from the relative in-plane displacement between the two layers. Because these registries favor different interlayer separations and orbital overlaps, they provide distinct local environments for low-energy states.

\begin{figure}[t]
  \centering
  \includegraphics[width=0.98\linewidth]{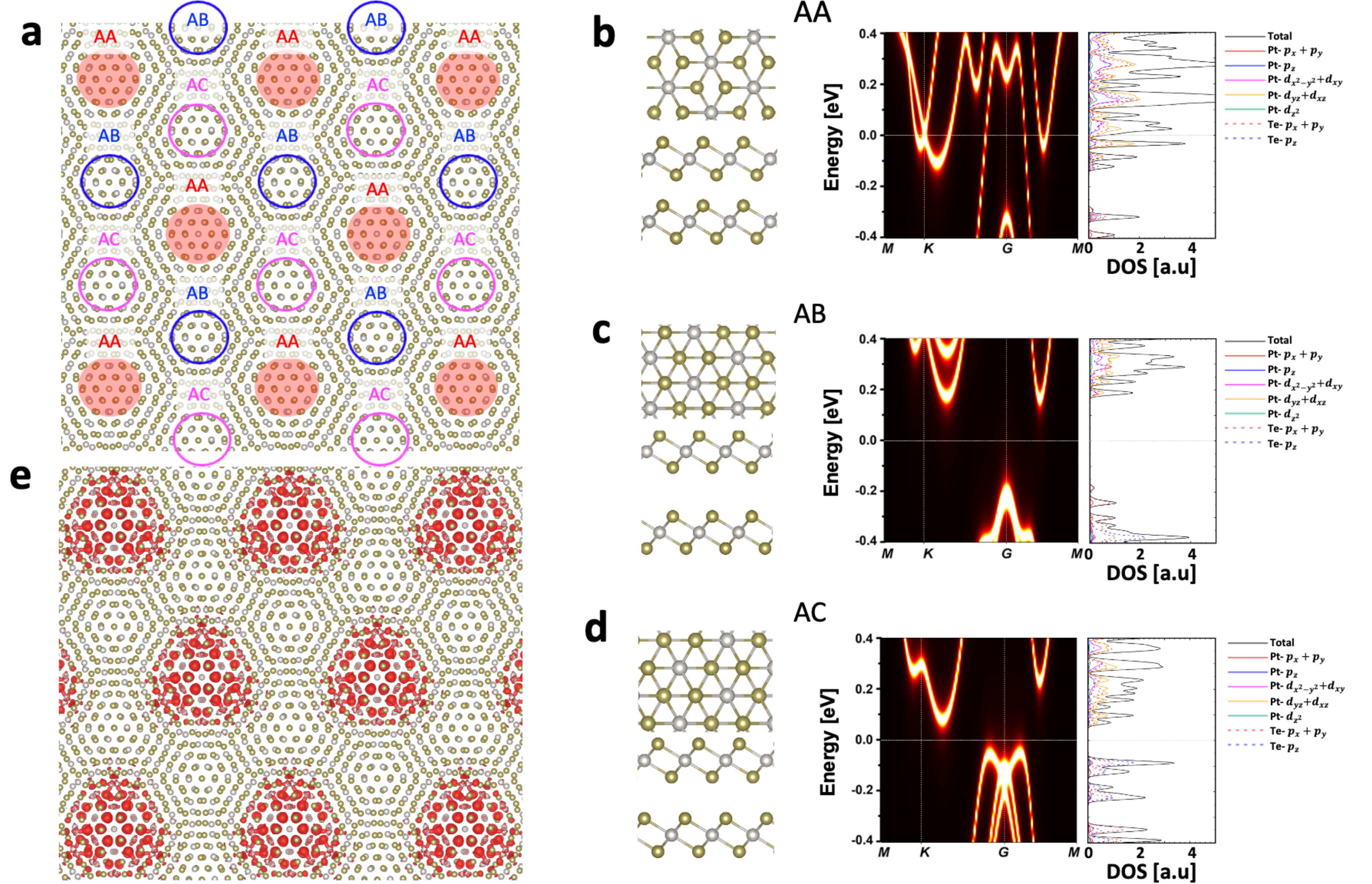}
  \caption{\textbf{Stacking-domain-dependent electronic character and low-energy localization.}
  \textbf{a,} Stacking-domain map of the $\theta=7.34^\circ$ moir\'e supercell, showing coexisting AA-like, AB-like, and AC-like regions. \textbf{b--d,} Reference calculations for ideal AA (\textbf{b}), AB (\textbf{c}), and AC (\textbf{d}) bilayers. The left panels show the corresponding stacking geometries, and the right panels show the unfolded spectra along $M$--$K$--$\Gamma$--$M$ and the orbital-projected densities of states. AA retains metallic spectral weight near $E_F$, whereas AB and AC show strong spectral suppression and/or a direct gap. \textbf{e,} Partial charge density of the selected low-energy states near $E_F$ in the $7.34^\circ$ structure. The low-energy weight is concentrated in AA-like regions, demonstrating domain-selective localization within the otherwise gapless moir\'e structure.}
  \label{fig3}
\end{figure}

Figure~\ref{fig3}a shows the stacking-domain map of the $7.34^\circ$ structure. Figures~\ref{fig3}b--d show separate reference calculations for bilayers with ideal AA, AB, and AC stackings. These reference spectra are not spatially resolved band structures of the full moir\'e supercell. Rather, they provide limiting local-registry configurations with which to interpret the moir\'e electronic structure. The AA reference remains metallic, while the AB and AC references show a strong reduction in spectral weight and/or a direct gap near $E_F$. The full $7.34^\circ$ moir\'e structure should therefore be described as a spatially inhomogeneous gapless state, not as a uniformly gapped semiconductor.

The real-space distribution of the near-$E_F$ states provides direct evidence for this electronic inhomogeneity. Figure~\ref{fig3}e shows the partial charge density of the selected low-energy states in the $7.34^\circ$ structure. The charge density is concentrated predominantly in the AA-like regions and is substantially reduced in the neighboring AB-like and AC-like regions. This spatial contrast demonstrates a pronounced domain dependence of the low-energy states within the otherwise gapless moir\'e cell. The ideal-registry calculations provide reference electronic structures consistent with this real-space distribution. Quantitative local band offsets, local spectral gaps, and transport across the domain boundaries require spatially resolved spectral functions or local densities of states and are not determined from the present partial charge density.

\subsection{Microscopic origin of the gap opening: separation-dependent Te-$p_z$ hybridization}\label{subsec:mechanism}

\begin{figure}[t]
  \centering
  \includegraphics[width=0.98\linewidth]{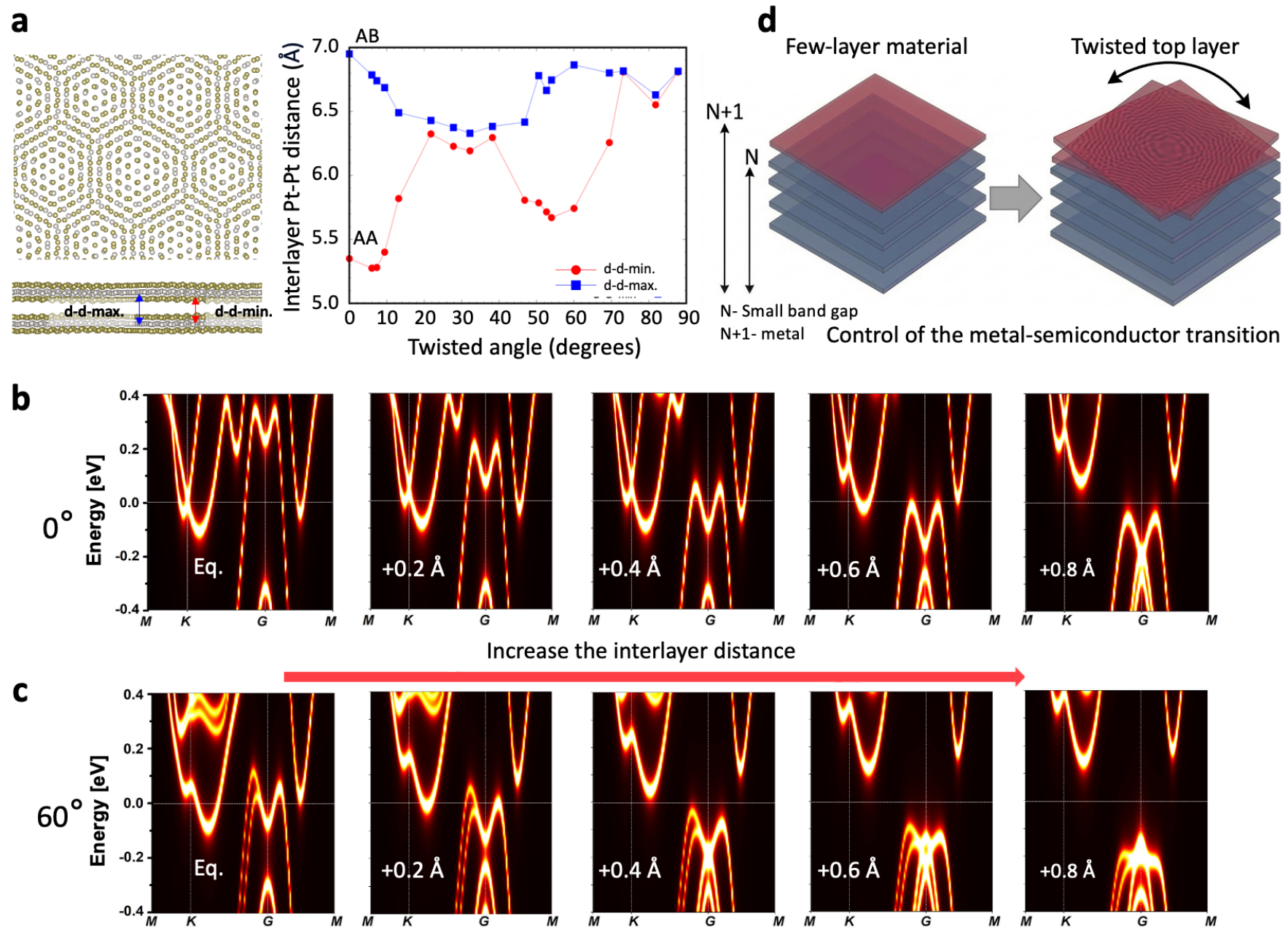}
  \caption{\textbf{Separation-dependent gap opening and interlayer hybridization.}
  \textbf{a,} Definition and twist-angle dependence of the minimum and maximum local interlayer Pt--Pt separations, $d_{\min}$ and $d_{\max}$. \textbf{b,} Controlled interlayer-separation scan for the $0^\circ$ bilayer at the relaxed equilibrium geometry (Eq.) and after increasing the separation by $+0.2$, $+0.4$, $+0.6$, and $+0.8$~\AA{}. \textbf{c,} Corresponding scan for the $60^\circ$ bilayer. In both cases, increasing the separation suppresses the near-$E_F$ crossings and opens a direct spectral gap. \textbf{d,} Schematic extension of the hybridization-controlled mechanism to layered materials whose low-energy bands are sensitive to interlayer coupling. In \textbf{b} and \textbf{c}, the in-plane geometry is fixed, thereby isolating the response to the out-of-plane separation within each stacking configuration.}
  \label{fig4}
\end{figure}

The correlation in Fig.~\ref{fig2}c suggests that the low-energy electronic phase is controlled by the local interlayer geometry. A twisted bilayer contains a distribution of local Pt--Pt separations rather than a single interlayer distance. We characterize this distribution by its minimum and maximum values, $d_{\min}$ and $d_{\max}$, as shown in Fig.~\ref{fig4}a. These quantities provide compact structural descriptors of the most strongly and weakly coupled local environments, although the electronic phase depends on the complete spatial distribution. In particular, the decrease of $d_{\min}$ in the sampled $60^\circ$ configuration coincides with the reappearance of spectral weight at $E_F$.

To isolate the effect of out-of-plane separation, we performed controlled interlayer-separation scans for the $0^\circ$ and $60^\circ$ bilayers. From each relaxed equilibrium structure, we increased the layer separation by $+0.2$, $+0.4$, $+0.6$, and $+0.8$~\AA{} while keeping the in-plane geometry fixed. In both structures, increasing the separation progressively suppresses the near-$E_F$ crossings and produces a finite direct gap (Fig.~\ref{fig4}b,c). Because the local in-plane registry is unchanged in these scans, the spectral evolution isolates the effect of the out-of-plane separation within the fixed stacking geometry.

This trend is consistent with the interlayer-hybridization mechanism identified previously for bilayer PtTe$_2$\cite{Kang2024PRR}. Increasing the layer separation reduces the interlayer Te-$p_z$ overlap and the associated bonding--antibonding splitting, thereby removing the band overlap near $E_F$. Twist controls the electronic phase by changing the spatial distribution of these local interlayer environments. The minimum separation captures the overall trend, while deviations from a one-parameter relation reflect the full registry and separation distribution of the moir\'e cell.

Figure~\ref{fig4}d summarizes a possible extension of this mechanism to other layered materials. In a layered material whose electronic phase is sensitive to the addition of one strongly coupled layer, twisting the top interface can redistribute the interlayer separation and thereby tune the low-energy hybridization. This mechanism may be relevant to other layered systems with strong thickness-dependent band-edge hybridization\cite{Chaves2020}, although its realization must be established separately for each material.

\section{Discussion}

Our calculations reveal a non-monotonic single-particle electronic phase sequence in twisted bilayer PtTe$_2$. The $7.34^\circ$ structure is gapless, with $E_g^{\mathrm{dir}}=0$, but its near-$E_F$ states are strongly localized in AA-like regions. Finite direct spectral gaps appear at the sampled intermediate angles; the gap closes at the sampled $60^\circ$ configuration and reopens at higher angles. The relevant control parameter is therefore not the moir\'e wavelength alone, but the registry-dependent distribution of interlayer environments generated by twist and structural relaxation.

The structural and electronic analyses consistently identify interlayer hybridization as the microscopic origin of the phase sequence. Larger local interlayer separations generally accompany larger direct gaps, and controlled separation scans at $0^\circ$ and $60^\circ$ show that increasing the separation is sufficient to remove the near-$E_F$ crossings within a fixed in-plane geometry. This behavior follows the established sensitivity of bilayer PtTe$_2$ to Te-$p_z$ hybridization and differs from a correlation-driven Mott mechanism.

The gapless $60^\circ$ result obtained here differs from the gapped configuration reported using PBE-D3 \cite{Li2025CPB}. Calculations performed at common geometries show that the $60^\circ$ structure lies close to an electronic boundary at which both the relaxed interlayer separation and the exchange--correlation treatment affect the band overlap. The nominal twist angle alone is therefore insufficient to specify the electronic regime of this near-critical configuration.

The combined unfolded spectra and Brillouin-zone-integrated densities of states establish the gapless or gapped character of the sampled structures within the numerical resolution of the calculations, while the unfolded spectra provide the reported direct-gap values. The $7.34^\circ$ structure additionally exhibits pronounced domain-selective localization of the near-$E_F$ states. Spatially resolved local densities of states, quantitative band offsets between local domains, and transport across domain boundaries remain important subjects for future work. Within these limits, the results demonstrate a single-particle route to moir\'e phase control in a semimetallic van der Waals bilayer through the redistribution of interlayer hybridization.

The predicted gap evolution and domain-dependent low-energy weight can be tested experimentally. Spatially resolved STM/STS should distinguish the enhanced near-$E_F$ spectral weight in the AA-like regions of the gapless $7.34^\circ$ structure from the reduced weight in neighboring domains. ARPES can test the disappearance and reappearance of near-$E_F$ crossings across the sampled twist angles, while gated transport can distinguish the gapless structures from configurations exhibiting a finite DOS gap. These measurements would provide complementary real-space, momentum-space, and transport tests of the calculated electronic evolution.

\section{Methods}\label{sec:methods}

\subsection{First-principles calculations and analysis of twisted bilayers}
First-principles calculations were carried out within density functional theory using the Vienna \textit{ab initio} simulation package (VASP)\cite{Kresse1993,Kresse1996}. The interactions between ionic cores and valence electrons were described using the projector-augmented-wave method\cite{PAW1994,Kresse1999}, and spin--orbit coupling was included throughout. Exchange--correlation effects were treated using the r$^2$SCAN meta-generalized-gradient approximation combined with the nonlocal van der Waals correction rVV10\cite{r2SCAN_rVV10}. In our previous benchmark study, r$^2$SCAN+rVV10 gave the closest overall agreement with diffusion Monte Carlo for stacking-dependent binding energies and equilibrium interlayer separations of bilayer PtTe$_2$\cite{Kang2024PRR}; it was used throughout this work. A plane-wave kinetic-energy cutoff of 400~eV was used. All structures were relaxed until the residual force on each atom was smaller than $5\times10^{-3}$~eV~\AA$^{-1}$, with the total energy converged to $10^{-5}$~eV. A vacuum region of at least 20~\AA{} was used to suppress interactions between periodic images.

Commensurate twisted bilayer supercells were constructed for selected twist angles between $0^\circ$ and $90^\circ$. For each structure, all atomic positions were relaxed, allowing both out-of-plane corrugation and in-plane reconstruction. Brillouin-zone integrations were performed using Monkhorst--Pack meshes\cite{monkhorst76}. A $15\times15\times1$ mesh was used for the untwisted bilayer. The meshes for the twisted supercells were reduced according to the supercell size to maintain a comparable reciprocal-space sampling density. Monolayer and bulk reference calculations used $12\times12\times1$ and $12\times12\times8$ meshes, respectively.

The supercell band structures were unfolded in the primitive-cell Brillouin zone and the corresponding spectral weights were evaluated along the $M$--$K$--$\Gamma$--$M$ path. The direct spectral gap $E_g^{\mathrm{dir}}$ was extracted from the separation between the highest occupied and lowest unoccupied unfolded states with appreciable spectral weight along this path. The phase assignment was cross-checked using the Brillouin-zone-integrated total density of states. A structure was classified as gapless when the unfolded spectral weight crossed $E_F$ and the total density of states remained finite at $E_F$; accordingly, the $7.34^\circ$ structure was assigned $E_g^{\mathrm{dir}}=0$. A structure was classified as gapped when a finite $E_g^{\mathrm{dir}}$ was resolved in the unfolded spectrum and the total density of states was suppressed to the numerical background level over a finite energy interval around $E_F$. The orbital-projected densities of states were used to identify the contributions of Pt-$d$ and Te-$p$ near the band edges.

The local stacking registry was characterized by the relative in-plane displacement between the two layers. Each real-space region was assigned to the nearest AA, AB, or AC high-symmetry registry in displacement space. This assignment was used to relate the relaxed local stacking pattern to the interlayer-separation distribution and to the real-space localization of the low-energy states. The local interlayer Pt--Pt separation was defined as the vertical distance between the Pt sublayers at the corresponding in-plane position.

For controlled interlayer-separation scans, the relaxed $0^\circ$ and $60^\circ$ structures were used as references. For each separation scan, one layer was translated rigidly along the out-of-plane direction. All intralayer coordinates, in-plane atomic positions, and lattice vectors were kept fixed, so that only the interlayer separation was changed. The resulting electronic structures therefore isolate the response to the layer separation within each fixed in-plane stacking geometry.



\section*{Acknowledgements}
S.-H. Kang was supported by Korea Institute of Science and Technology Information (KISTI) (K26L4M2C2-01).
Y.-K. Kwon were supported by the National Research Foundation (NRF) of Korea grants funded by the Korean government (MSIT) (No. RS-2025-25457100 and No. RS-2024-00416976). J. Ahn acknowledges support from the U.S. Department of Energy, Office of Science, Office of Basic Energy Sciences, under Computational Materials Sciences Award No.~DE-SC0020177. 

\section*{Author contributions}
S.-H.K. conceived the project and performed the calculations. S.-H.K. and J.A. analyzed the results. S.-H.K. wrote the first draft. J.A. and Y.-K.K. revised the manuscript. All authors discussed the results and contributed to the final manuscript.


\bibliography{biblio}

\end{document}